\documentclass[aps,prl,letterpaper,amssymb,twocolumn,preprintnumbers,floatfix,superscriptaddress]{revtex4}
\usepackage{graphicx}
\usepackage{dcolumn}
\usepackage{bm}
\usepackage{hyperref}
\usepackage{epstopdf}
\usepackage{color}
\usepackage{amsmath}

\def\lsim{\raisebox{-4pt}{$\,\stackrel{\textstyle{<}}{\sim}\,$}}
\def\gsim{\raisebox{-4pt}{$\,\stackrel{\textstyle{>}}{\sim}\,$}}

\def\beq{\begin{equation}} 
\def\eeq{\end{equation}} 
\def\be{\begin{equation}} 
\def\ee{\end{equation}} 
\def\bea{\begin{eqnarray}} 
\def\eea{\end{eqnarray}} 
\def\ben{\begin{enumerate}} 
\def\een{\end{enumerate}}

\def\lsim{\mathrel{\raise.3ex\hbox{$<$\kern-.75em\lower1ex\hbox{$\sim$}}}} 
\def\gsim{\mathrel{\raise.3ex\hbox{$>$\kern-.75em\lower1ex\hbox{$\sim$}}}} 
\def\ifmath#1{\relax\ifmmode #1\else $#1$\fi}

\begin{document}
\DeclareGraphicsExtensions{.jpg,.pdf,.mps,.png,}

\title{Low Scale Composite Higgs Model and 1.8 $\sim$ 2 TeV Diboson Excess }


\author{Ligong Bian}
\affiliation{State Key Laboratory of Theoretical Physics and Kavli Institute for Theoretical Physics China, Institute of Theoretical Physics, Chinese Academy of Sciences, Beijing 100190, P. R. China.}
\author{Da Liu}
\affiliation{State Key Laboratory of Theoretical Physics and Kavli Institute for Theoretical Physics China, Institute of Theoretical Physics, Chinese Academy of Sciences, Beijing 100190, P. R. China.}
\author{Jing Shu}
\email{e-mail: jshu@itp.ac.cn}
\affiliation{State Key Laboratory of Theoretical Physics and Kavli Institute for Theoretical Physics China, Institute of Theoretical Physics, Chinese Academy of Sciences, Beijing 100190, P. R. China.}
\affiliation{CAS Center for Excellence in Particle Physics, Beijing 100049, China.}

\begin{abstract}

We consider a simple solution to explain the recent diboson excess observed by ATALS and CMS Collaborations in models with custodial symmetry $SU(2)_L \times SU(2)_R \rightarrow SU(2)_c$. The $SU(2)_L$ triplet vector boson $\rho$ with mass range of $1.8 \sim 2$ TeV would be produced through the Drell-Yan process with sizable diboson decay branching to account for the excess. The other $SU(2)_L \times SU(2)_R$ bidoublet axial vector boson $a$ would cancel all deviations of electroweak obervables induced by $\rho$ even if the SM fermions mix with some heavy vector like (composite) fermions which couple to $\rho$ (``non-universally partially composite"), therefore allows arbitrary couplings between each SM fermion and $\rho$. We present our model in the ``General Composite Higgs" framework with $SO(5) \times U(1)_X \rightarrow SO(4) \times U(1)_X$ breaking at scale $f$ and demand the first Weinberg sum rule and positive gauge boson form factors as the theoretical constraints. We find that our model can fit the diboson excess very well if the left-handed SM light quarks, charged leptons and tops have zero, zero/moderately small and moderate/large composite components for reasonable values of $g_\rho$ and $f$. The correlation between tree level $S$ parameter and the $h \rightarrow Z \gamma$ suggest a large $a$ contribution to $h \rightarrow Z \gamma$ and it is indeed a $\mathcal{O}(1)$ effect in our parameter space which provides a strong hint for our scenario if this diboson excess is confirmed by the $13 \sim 14$ TeV LHC Run II.
\end{abstract}

\maketitle

\section{Introduction}

The Standard Model (SM) of particle physics, which is our current deepest understanding of microscopic physics, has been intensively tested in the last few decades. At present, we are in a stage to probe new physics at the Terascale by the Large Hadron physics at CERN. All scientists in particle physics are looking for the new data and see if there are some hints for physics beyond SM. 

Very recently, the ATALS and CMS collaboration have reported several excesses over the SM backgrounds in the invariant mass distributions around 1.8 $\sim$ 2 TeV. (1): the ATLAS collaboration~\cite{Aad:2015owa} has reported that a 3.4$\sigma$, 2.6$\sigma$ and 2.9$\sigma$ deviation are observed around 2 TeV in the invariant mass distribution of boosted WZ, WW and ZZ \footnote{Given that the mass resolution in the jet mass reconstruction of the W and Z is $\pm 13$ GeV, it is logically possible that the ZZ reconstructed events do in fact involve Ws.}, with the global significance of the discrepancy in the WZ channel being 2.5$\sigma$. The CMS experiment reported a moderate excess, about $1.4\sigma$ for the dijet resonances, without distinguishing between the W- and Z-tagged jets ~\cite{Khachatryan:2014hpa,Khachatryan:2014gha}. (2): CMS experiment reported a $\sim$2$\sigma$ excess slightly below 2 TeV($\sim$ 1.8 TeV) in the dijet resonance channel search~\cite{Khachatryan:2015sja} and $W^\prime\rightarrow Wh \rightarrow b\bar{b} l \nu$ search with a highly boosted SM like Higgs.

To explain the above excesses, the new physics contributions to the resonance production of $WZ$, $WW$ and $ZZ$ are required to be around 6 $\sim$ 10 fb. There have been several papers on explaining this diboson excesses at 1.8 $\sim$ 2 TeV \cite{Cacciapaglia:2015nga, Carmona:2015xaa, Gao:2015irw, Cao:2015lia, Cheung:2015nha, Abe:2015uaa, Sanz:2015zha, Thamm:2015csa, Anchordoqui:2015uea, Omura:2015nwa, Chen:2015xql, Chiang:2015lqa, Dobrescu:2015yba, Brehmer:2015cia,Dobrescu:2015qna,Hisano:2015gna,Chao:2015eea} by either some massive spin one vector bosons ($\rho$) or exotic spin zero or two objects. For spin one explanations, the vector bosons are either based on $SU(2)_1 \times SU(2)_2 \times U(1)_X \rightarrow SU(2)_L \times U(1)_Y$ or $SU(2)_L \times SU(2)_R \times U(1)_{B-L} \rightarrow SU(2)_L \times U(1)_Y$ symmetry breaking pattern or some extended variations. The former, which is the prototype of many models explaining electroweak symmetry breaking (EWSB), in general induce a large tree-level $S$ parameter $S = 4 \pi v^2/ m_{\rho}^2 = 0.2$ (in gauge eigenstate basis) through the kinetic term of Higgs and scalars which triggers the symmetry breaking \footnote{The models based on the latter symmetry breaking pattern are less constrained from EWPT \cite{Shu:2011au}.}. While this large positive $S$ alone might marginally fit the $S-T$ plane with a very optimal positive $T$, other effects from one-loop diagrams of heavy resonances~\cite{Orgogozo:2012ct, Contino:2015mha}, modified Higgs couplings~\cite{Barbieri:2007bh} and effective operators from UV~\cite{Contino:2011np} would spoil the electroweak precision tests (EWPT). The other difficulty is that while a large $q\bar{q} \rho$ coupling is required to explain the diboson excess, constraints from $pp \rightarrow \rho^0 \rightarrow l^+ l^-$ would violate fermion universality in the $f \bar{f} \rho$ couplings which further contributes to EWPT \footnote{This problem is even more severe if there are other large decay channels of $\rho$. for instances, in composite Higgs models where a large $t \bar{t} \rho^0$ or $t \bar{b} \rho^+$ coupling is expected.}.


Here we consider a simple solution of the above two issues to account for the diboson excess by introducing an axial vector boson $a$ which transforms as $(2,2)$ representation of $SU(2)_L \times SU(2)_R$ which do not couples to the SM fermions before EWSB. For convenience, we use the ``General Composite Higgs" set up of Ref. \cite{Marzocca:2012zn} which is the CCWZ construction of $SO(5)/SO(4) \times U(1)_X$ with vector meson dominance for spin one fields as an illustration. We find that the $a$ field contributions can not only cancel the large positive $S$, but also the couplings between composite fermions and SM gauge bosons by integrating out $\rho$. The latter allows the partially composite SM fermions to have tiny EWPT contributes with arbitrary couplings to $\rho$ in a very wide range. In our model of vector bosons ($\rho$ and $a$), we use the first Weinberg's sum rule and positive EWSB form factor  (in the spirit of Witten's theorem \cite{Witten:1983ut}) as our theoretical constraints. We find that the current diboson excess can be explained very well for $\xi \sim (0.05, 0.2)$, $g_\rho \sim (1.5, 2.5) $ and left-handed SM light quarks, charged leptons and top/bottoms have zero, zero or moderately small and moderate or large composite components composite components where the last one is a good sign for correct Coleman Weinberg Higgs potential. Notice that the $S$ parameter and $h \rightarrow Z \gamma$ decay can come from the same operators which has large correlations, we also calculate the $a$ contribution to $h \rightarrow Z \gamma$ and find that the corrections are very large ($\mathcal{O}(1)$ of the SM one). 

\section{Basic setup of composite Higgs Model}

In the Minimal Composite Higgs Model, the Higgs arises as a pseudo-Nambu
Goldstone boson (pNGB) from the spontaneous symmetry breaking of the
global symmetry group $SO(5)\times U(1)_{X}\rightarrow SO(4)\times U(1)_{X}$
of the strongly-interacting sector. The unbroken $SO(4)$ is isomorphic to $SU(2)_L \times SU(2)_R $, so the custodial symmetry is preserved. The extra $U(1)_X$ is introduced to reproduce the right hyper-charge for the SM fermions. Since it doesn't play an important role in our following discussion, we will neglect it from now on. The four NGBs $h^{\hat{a}}$ can
be described as the fluctuations along the broken generators $T^{\hat{a}}$
($\hat{a}=1,\ldots,4$): $U = \mbox{exp} (i \sqrt 2 \pi^{\hat a} T^{\hat a}/f)$, where $f$ is the decay constant for those pNGB fields. Under a transformation $g\in SO(5)$, this field transforms non-linearly
as $U\rightarrow gUh^{\dagger}\left(g,h^{\hat{a}}(x)\right)$, where
$h\in SO(4)$. Going to the unitary gauge, we can
bring the NGBs to the form $h^{\hat{a}}=(0,0,0,h)$. With this choice,
the matrix $U$ takes the simple form\begin{equation}
U=\left(\begin{array}{ccccc}
1 & 0 & 0 & 0 & 0\\
0 & 1 & 0 & 0 & 0\\
0 & 0 & 1 & 0 & 0\\
0 & 0 & 0 & \cos\frac{h}{f} & -\sin\frac{h}{f}\\
0 & 0 & 0 & \sin\frac{h}{f} & \cos\frac{h}{f}\end{array}\right).\label{Uunitary}
\end{equation}
The interactions with the elementary fields of the SM explicitly break
the $SO(5)$ symmetry and will induce a potential for the Higgs through
loop corrections, such that modulus of the NGB fields, $h$, acquires
a vev breaking $SO(4)$ to the custodial subgroup $SO(3)_{c}$ and
the electroweak symmetry to $U(1)_{\text{em}}$, at the scale $v=f\sqrt{\xi}=f\sin\langle h\rangle/f$. The CCWZ structures \cite{CCWZ} are defined by $i U^{\dagger}D_{\mu}U=d_{\mu}^{\hat{a}}T^{\hat{a}}+E_{\mu}^{a}T^{a}$ with  $T^{\hat a } \, , \hat a = 1,2,3,4$ being generators in the broken direction and $T^{a_L,a_R } \,, a_{L,R} = 1,2, 3$ being generators in the unbroken direction. It follows that  $d_\mu$ transforms covariantly under the local symmetry group, while  $E_\mu$ transforms like a gauge field. Expanding over $h/f$, we have:
\begin{equation}
\begin{cases}
d_{\mu}^{\hat{a}}= & - \frac{\sqrt{2}}{f}(D_{\mu}h)^{\hat{a}}+\ldots\\
E_{\mu}^{a}= & g_{0}A_{\mu}^{a} + \frac{i}{f^{2}}(h\stackrel{\leftrightarrow}{D_{\mu}}h)^{a}+\ldots\end{cases}\end{equation}

At the leading order in the chiral expansion, the Lagrangian describing
the dynamics of the NGBs is given by\begin{equation}
\mathcal{L}=\frac{f^{2}}{4}\text{Tr}\left(d_{\mu}d^{\mu}\right)=\frac{f^{2}}{2}(\partial_{\mu}\Sigma)^{t}(\partial^{\mu}\Sigma),\end{equation}
where we have defined a linearly-tranformed field by using $SO(4)$ invariant vacuum $\Sigma_{0}^{t}=(0,0,0,0,1)$:
\beq
\Sigma=U(h^{\hat{a}})\Sigma_{0}=\frac{\sin h/f}{h}(h^{1},h^{2},h^{3},h^{4},h\cot h/f).
\eeq

The gauging of the EW symmetry group $SU(2)_{L}\times U(1)_{Y}$ can
be performed in the chiral Lagrangian showed above just by promoting
the derivative to a covariant one, $\partial_{\mu}\rightarrow\partial_{\mu}- i(g_{0}W_{\mu}^{a}T_{L}^{a}+g_{0}^{\prime}B_{\mu}T_{R}^{3})$.
The leading order Lagrangian for the gauge fields and the Nambu-Goldstone
bosons then reads\begin{equation}
\mathcal{L}^{(0)}=-\frac{1}{4}W_{\mu\nu}^{a}W^{a\mu\nu}-\frac{1}{4}B_{\mu\nu}B^{\mu\nu}+\frac{f^{2}}{2}(D_{\mu}\Sigma)^{t}(D^{\mu}\Sigma).\end{equation}
To give formal $SO(5)$ transformations to the gauge bosons, one can
use the method of spurions introducing fake fields to complete the
$SO(5)$ multiplet and gauge the whole group: $A_{\mu}=A_{\mu}^{a}T^{a}+A_{\mu}^{\hat{a}}T^{\hat{a}}$. 

The most general $SO(5)\times U(1)_{X}$-invariant Lagrangian depending
on the gauge fields and the NGBs, at the quadratic order in the gauge
fields and in momentum space, is
\beq
\mathcal{L}^{eff} = \frac{P_{T}^{\mu\nu}}{2}(\Pi_{0}(p^{2})\text{Tr}(A_{\mu}A_{\nu})  +\Pi_{1}(p^{2})\Sigma^{t}A_{\mu}A_{\nu}\Sigma).
\label{eq:generalSO5}
\eeq
where $P_{T}^{\mu\nu}=\eta^{\mu\nu}-p^{\mu}p^{\nu}/p^{2}$ is the
projector on the transverse field configurations. Performing an $SO(4)$
transformation to get $\vec{h}=(0,0,0,h)$ and turning off the spurionic
gauge fields keeping only the $SU(2)_{L}\times U(1)_{Y}$ ones, namely $A_{\mu}^{a_{L}}=W_{\mu}^{a}$, $A_{\mu}^{3_{R}}= B_{\mu} $.


The effective Lagrangian for the SM gauge bosons in $SO(5)/SO(4)$ with the explicit dependence on the Higgs field can be written as:
\bea
\mathcal{L}^{eff}= & \frac{P_{T}^{\mu\nu}}{2}  \bigg( \Pi_{0}W_{\mu}^{a}W_{\nu}^{a} + \Pi_{1} \frac{s_h^2}{4}\left(W_{\mu}^{1}W_{\nu}^{1} + W_{\mu}^{2}W_{\nu}^{2}\right) +  \nonumber \\
 & \Pi_{B} B_{\mu}B_{\nu} + \Pi_{1} \frac{s_h^2}{4}\left(\frac{g_0^\prime}{g_0} B_{\mu}-W_{\mu}^{3}\right)\left(\frac{g_0^\prime}{g_0} B_{\nu}-W_{\nu}^{3}\right) \nonumber \\
 &  + c_h \Pi_{LR} \left( W_{\mu}^{a}W_{\nu}^{a} - \frac{g_0^{\prime 2}}{g_0^2}B_\mu B_\nu \right)  \bigg),
\eea
where $\Pi_B =  g_0^{\prime 2}/g_0^2 \Pi_0$. From this Lagrangian one can easily obtain the relation $\Pi_{W_3B} =-g_0^{\prime }/g_0 \frac{s_h^2}4\Pi_1$, which is directly related with S-parameter.


\section{The spin one resonances}


In this paper, we are going to introduce one vector resonance
$\rho^{a_L, a_R}_\mu$ transforming as $(3,1)\oplus (1,3)$
and one axial  resonance $a_\mu^{\hat a}$  transforming as
$(2,2)$  under the $SU(2)_L \times SU(2)_R$  symmetry group.  The
Lagrangian for vector and  axial resonances could be summarized by
the following equations~\cite{Marzocca}:
\bea
\mathcal{L}_{\rho_L} &=&  - \frac{1}{4} \mbox{Tr} \left( \rho _{L,\mu \nu } \rho_L^{\mu \nu } \right) + \frac{ f_\rho^2}{2}  \mbox{Tr} {\left( g_{\rho} \rho_{L\mu} - E_\mu ^L \right)^2}   \label{lr} \\
 \mathcal{L}_{\rho_R}  &=&  - \frac{1}{4} \mbox{Tr} \left( \rho_{R,\mu \nu } \rho_R^{\mu \nu } \right) + \frac{f_\rho^2}{2} \mbox{Tr}  {\left( g_{\rho} \rho_{R\mu} - E_\mu ^R\right)^2}   \label{rr}  \\
\mathcal{L}_a  &=&  - \frac{1}{4} \mbox{Tr} \left( a_{\mu \nu } a^{\mu \nu } \right) + \frac{f_a^2}{2 \, \Delta^2}  \mbox{Tr} {( {g_a} {a_\mu } - \Delta  {d_\mu } )^2}  ,  
\eea
where the mass parameters can be defined as: $m_\rho =  g_\rho f_\rho$, $ m_a = g_a f_a /|\Delta|$. 
The electroweak field strengthes are: $ W_{\mu \nu} = \partial_\mu W_\nu -\partial_\nu W_\mu - i g_0 \left[ W_\mu ,  W_\nu \right]$, $B_{\mu \nu} = \partial_\mu B_\nu -\partial_\nu B_\mu$.  For the  axial resonances, the notation in the kinetic term reads:
\bea
a_{\mu \nu } = \nabla _\mu a_\nu  - \nabla _\nu a_\mu \, ,  \quad  \nabla _\mu  = \partial _\mu  - i E_\mu  \nonumber
\eea
The leading tri-linear interaction terms involving one Higgs boson read:
\begin{equation}
\label{Lag}
\begin{split}
\mathcal{L}_h = & 
(\frac14 f^2  + \frac12 f_a^2)\sqrt{\xi}\sqrt{1-\xi} \frac{h}{f} (g_0 W_{\mu}^a - g_0^\prime \delta^{a3} B_\mu)^2\\
& +\frac{f_{\rho_L}^2}{2} \sqrt{\xi} \frac{h}{f}
\left( g_{\rho_L} \rho_\mu^{aL} - g_0 W_\mu^a)(g_0 W_{\mu}^a - g_0^\prime \delta^{a3} B_\mu) \right)\\
&-\frac{f_{\rho_R}^2}{2} \sqrt{\xi} \frac{h}{f}
\left( g_{\rho_R} \rho_\mu^{aR} - g_0^\prime \delta^{a3} B_\mu)(g_0 W_{\mu}^a - g_0^\prime \delta^{a3} B_\mu \right)\\
&-\frac{f_a^2}{\sqrt{2} \Delta} \sqrt{1 - \xi} \frac{h}{f}
 g_a a_\mu^a (g_0 W_{\mu}^a - g_0^\prime \delta^{a3} B_\mu) \\
\end{split}
\end{equation}
with $\xi = v^2/f^2 $,  and  $ a = 1,2, 3 $. Note that $ h\rho W$ coupling is suppressed by an additional $\sqrt{\xi}$ compared with that of the axial resonance and as a result, its contribution to the Higgs process is subdominant. See  below for the calculation of the $h\rightarrow Z \gamma$ decay width.


\section{ The S parameter and electroweak precision constraints} 

The explicit form of the form factors is obtained by integrating out the heavy vector resonances at tree-level. 
For simplicity, assuming only one vector and axial resonances, we simply get
 \bea
\Pi_{1}(p^{2})&= & g_{0}^{2}f^{2}+2 g_{0}^{2}p^{2}\left[ \frac{f_{a}^{2}}{(p^{2} + m_{a}^{2})}-\frac{f_{\rho}^{2}}{(p^{2} + m_{\rho}^{2})}\right],\\
\Pi_{0}(p^{2})&= & p^{2}+g_{0}^{2}p^{2} \frac{f_{\rho}^{2}}{(p^{2}+m_{\rho}^{2})},\ \ \Pi_0^X(p^2) = p^2\,.
\label{PiFF}
\eea


The $\Pi_{1}(p^{2})$ form factor involves NGBs and measures EWSB effects. We regulate its UV behavior by imposing $\Pi_{1}(p^{2})$ to zero at high energy as the first Weinberg sum rule in QCD (\cite{Weinberg:1967kj}):
\begin{equation}
\lim_{p^{2}\rightarrow + \infty} g_0^{-2} \Pi_{1}(p^{2})=f^{2}+2 f_{a}^{2} -2 f_{\rho}^{2} \equiv0\,. \ \ \ ({\rm I})
\label{SRI} 
\end{equation}

Unlike the QCD,  we do not impose $\Pi_{1}$ goes to zero faster than $1/p^{2}$ for large momenta. Instead, we use a parameter $\alpha$ categorizes the deviation to relaxing the second Weinberg's sum rule, we have 
\begin{equation}
\lim_{p^{2}\rightarrow + \infty}g_0^{-2} p^{2}\Pi_{1}(p^{2}) = 2 (f_{\rho}^{2} m_{\rho}^2 - f_{a}^{2} m_{a}^2)
= 2 \alpha^2 f^4.
\label{SRIInew}
\end{equation}
where the natural size for the $\alpha$ is $\alpha \sim g_\rho, g_a$.



The form factor $\Pi_{W_3 B}$ is also related to the $S$-parameter:
\be
S =  - \frac{16\pi}{g g'} \Pi_{W_3B}^\prime(0) = \frac{16\pi}{g^2}\frac{s_h^2}{4} \Pi^\prime_1(0)
\label{Spara}
\ee
As well known, $S$ is the main phenomenological electroweak bound constraining Composite Higgs Models, that requires $s_h\ll1$. From eq.\eqref{PiFF} and eq.\eqref{Spara}, we readily obtain the tree-level contribution to the $S$-parameter:
\bea
S \simeq 8 \pi s_h^2 \left( \frac{f_{\rho}^{2}}{m_{\rho}^2}- \frac{ f_{a}^{2}}{ m_{a}^2} \right) 
\label{SparaGauge}
\eea
where we have approximated $g_0\simeq g$ for simplicity. \footnote{It is important to notice that all the calculations in this section are based on the gauge eigenstate, where in many other papers like Ref. \cite{Carmona:2015xaa}, the EW precision observables are calculated in the Kaluza-Klein basis which is the mass eigenstate basis before EWSB. In this case, the tree level $S$ parameter is transmitted into universal shift between SM gauge bosons and light fermions \cite{Grojean, Agashe:2003zs}, where additional anomalous triple gauge boson couplings are also expected and might be probed in the future \cite{Bian:2015zha}.}. In the simple case without $a$, we have $S = {4\pi v^2}/{m_\rho^2}$ which marginally fits $S \lesssim 0.25$ when $m_{\rho} \gtrsim 2$ TeV. Notice that in our parametrization, we do not expect $m_\rho \simeq m_a$ and $f_\rho \simeq f_a$ in the ``walking region"  to get a suppressed tree level $S$ parameter.

It is interesting to think about the generic condition $\Pi_1(p^2) \geqslant 0$ as suggested by Witten's theorem if the underlying strong dynamics is vector confining \cite{Witten:1983ut}. In this case, the first Weinberg sum rule Eq. (\ref{SRI}) suggests that $\Pi_1(0) \geqslant 0$ since $f_\rho  \geqslant f_a$ and Eq. (\ref{SRIInew}) guarantees that $\Pi_1( + \infty)  \geqslant 0$. There could be one other minima $\Pi_1(p_*^2) = (f_a m_a + f_\rho m_\rho)^2/(m_\rho^2 - m_a^2)$ in the intermediate region ($\Pi'_1(p_*^2) = 0$)  $p_*^2 = m_a m_\rho (f_a m_\rho - f_\rho m_a)/ (f_\rho m_\rho - f_a m_a) \geqslant 0$ if $S<0$, which suggests that $m_\rho  \geqslant m_a$ for negative $S$. Therefore in vector confining theory, $S$ is always positive if $m_\rho < m_a$. 

There are also contributions from the composite Higgs at one-loop:
\bea
\Delta {S} &=& \frac{1}{6 \pi} \left[ s_h^2 \log \left( \frac{\Lambda }{m_h} \right) + \log \left( \frac{m_h}{m_{h,ref}}\right)\right]   \,  ,    \\
\Delta {T} &=&  - \frac{3  }{8 \pi c_w^2} \left[ s_h^2 \log \left( \frac{\Lambda }{m_h} \right)   + \log \left( \frac{m_h}{m_{h,ref}}\right) \right]
\label{IR} 
\eea



\begin{figure}[t]
\includegraphics[width=.99\columnwidth]{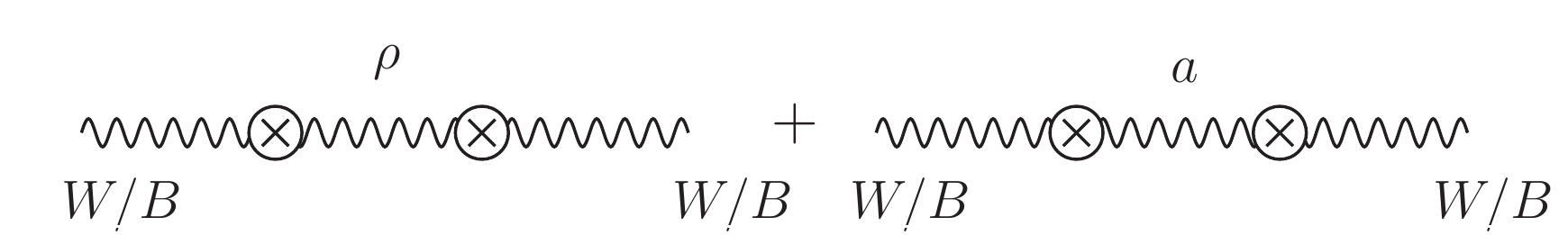} \\
\includegraphics[width=.99\columnwidth]{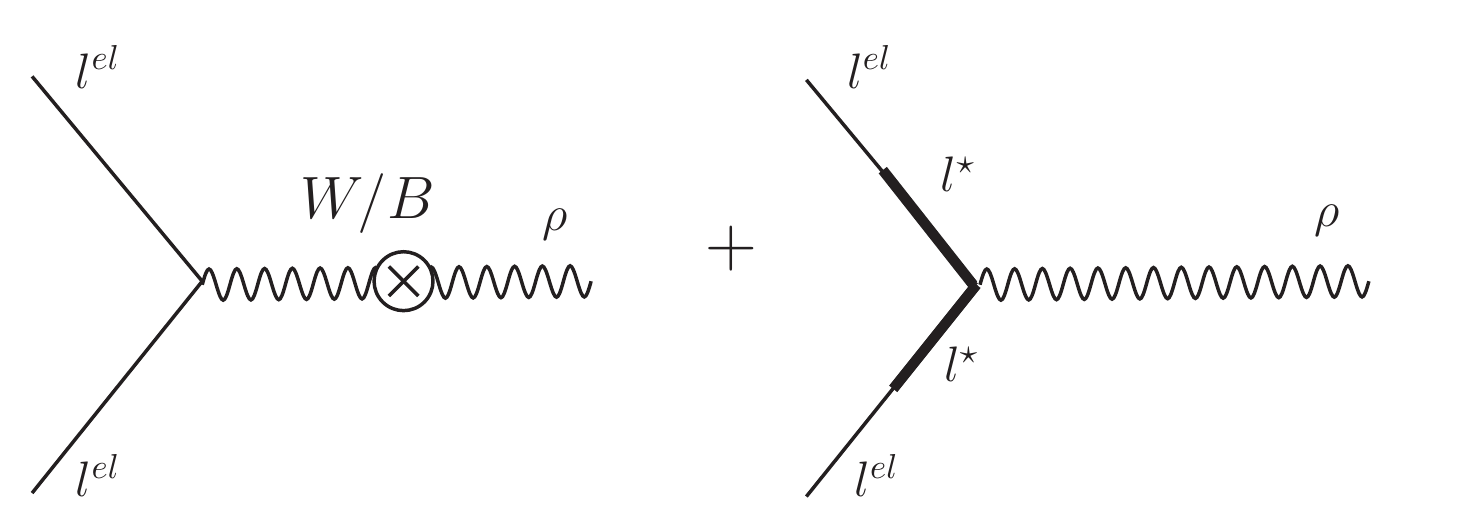} \\ 
\includegraphics[width=.99\columnwidth]{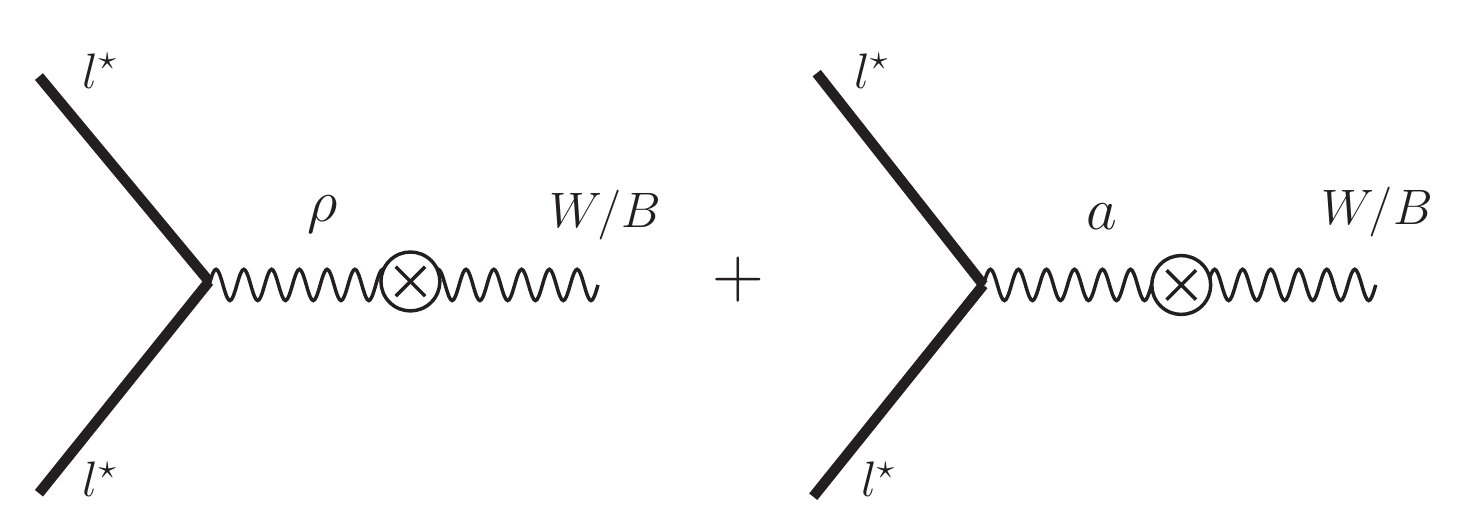}
\caption{Cancellation mechanisms between the $\rho$ and $a$ contributions to the EW precision observables and $\rho-l-l$ couplings. Upper: cancellation occurs between the contributions to the SM gauge boson vacuum polarization diagram. 
Middle: cancellation occurs between the $\rho$ and SM lepton couplings.. 
Bottom: cancellation occurs between the SM gauge boson and composite fermion couplings.}
\label{EWdia}
\end{figure}



%


 
 
In order to decrease the lepton $\rho$ coupling which could make our model less constrained by the $\rho \rightarrow l^+ l^-$ decay, we allow our leptons mix moderately with composite leptons $l^*$ which is of order $g/\sqrt{g_\rho g_\rho^{l*}}$, where $g_\rho^{l*}$ is the coupling between $\rho$ and composite leptons (See the middle panel of Fig.\ref{EWdia}). On the other hand, this extra composite components of light fermions will induce operators like i$(l^+ \gamma^\mu l)(H^\dag D_\mu H)$ by integrating out the vector fields $\rho$ and $a$. Because of the existence of $a$, one can adjust the coupling between $a$ and $l^*$ ($g_\rho^{l*} (\xi/2) m_\rho^2 /g_\rho \sim g_a^{l*} \sqrt{\xi/2}m_a^2 g_a $) so that couplings between $l^*$ and SM gauge bosons $W/B$ are further suppressed or even cancelled \footnote{Notice that our cancellation mechanism here is very different from the ``delocalization" scenario in Higgsless theory \cite{Csaki:2003zu, Cacciapaglia:2004rb}. In the ``delocalization" scenario, cancellation occurs between elementary and composite components of SM light fermions when integrating the $\rho$, so that the localization or the elementary-composite mixture of SM fermions is fixed for a given model of $\rho$. While in our scenario, cancellation occurs in the coupling of composite fermions and SM gauge bosons between integrating out $\rho$ and $a$, therefore allows arbitrary localization or elementary-composite mixture.  }.

\section{Fit the LHC 2TeV diboson excess}

The ATLAS (\cite{Aad:2015owa}) and CMS (\cite{CMSWZhad_2014}) collaborations have reported the observed and expected bounds at 95\% C.L. around 2 TeV in the 
diboson invariant mass distributions, which we summarize in  
 Table.~\ref{excess}. We can see clearly the excess  in both CMS and ATLAS.  Note that the search is dedicated to the  hadronically decaying Z and W  and the ability to accurately distinguish the two gauge bosons is highly limited,  the three channels are not exclusive. 
To explain the excesses with new physics, the new physics
need to contribute somehow $6-8$ fb~\cite{Carmona:2015xaa}.  

In this section, we focus on the  $\rho_L$ and investigate the possibility to account for the excesses of $\sigma$(WZ) and $\sigma$(WW) at ATLAS (\cite{Thamm:2015csa,Cao:2015lia,Abe:2015uaa,
Cheung:2015nha,Sanz:2015zha,Gao:2015irw,Carmona:2015xaa,
Cacciapaglia:2015nga}) \footnote{For $\rho_R$, the couplings with SM fermions are suppressed by an additional $\xi$ for the charged resonance and $g^{\prime 2}/g^2$ for the neutral resonance. It may not account for the $WZ$ excess even including the contamination from $WW$ channel.}.
The two parameters ($m_a, f_a$) of the axial resonance can be obtained by  the Weinberg first and relaxed second sum rule (eq.~\ref{SRI}, \ref{SRIInew}) as functions of $(\xi, m_\rho, g_\rho, \alpha)$. All right-handed SM model fermions are not relevant here since we embed them in the singlet representation of $SO(4)$. For SM top quark, only the compositeness of $t_L$ plays an important role in the phenomenology of $\rho_L$. Here we introduce the parameter $\epsilon_{tL}$ to measure the degree of compositeness for $t_L$ (see appendix for detail).
Now we have five  parameters in our model $(M_\rho, g_\rho, \xi, \alpha, \epsilon_{tL})$, where $M^2_{\rho} \sim m^2_\rho (1 + g^2/g_\rho^2)$ is the physical mass of the $\rho^{\pm,0}$. We will  work in the large coupling limit $g_{\rho} \gg g, g_0 \sim g, g_0^\prime \sim g^\prime$  and at leading order in $\xi$. The convention and expressions for the couplings can be found in the appendix.

\begin{table}[htb]
\begin{center}
\begin{tabular}{|lcr|c|c|c|c|}
	\hline
\multicolumn{3}{|c|}{Chanel} & 1.8 TeV & 2.0 TeV\\
\hline
ATLAS&WZ &\cite{Aad:2015owa} & 30 fb (14 fb) &38 fb (11 fb)\\
ATLAS&WW&\cite{Aad:2015owa}& 19 fb (19 fb)& 30 fb (14 fb)\\
ATLAS&ZZ&\cite{Aad:2015owa}& 29 fb (11 fb)& 30 fb (10 fb) \\
CMS&WZ &\cite{CMSWZhad_2014} & 17 fb (12 fb) &14 fb (8 fb)\\
CMS&WW&\cite{CMSWZhad_2014}&38 fb (29 fb) & 28 fb (20 fb)\\
CMS&ZZ&\cite{CMSWZhad_2014}& 28 fb (21 fb)& 23 fb (14 fb) \\
CMS&WH&\cite{Khachatryan:2015bma}& 15 fb (9 fb)& 7 fb (7 fb) \\
CMS&ZH&\cite{Khachatryan:2015bma}& 12 fb (8 fb)& 7 fb (7 fb) \\
\hline
\end{tabular}
\end{center}
\caption{Summary of the observed and expected bounds at 95\% C.L. around 2 TeV at ATLAS and CMS, with and without the brace to denote expected and observed values. }
\label{excess}
\end{table}

\section{Explanation of the diboson excess and constraints from the LHC}

\begin{figure}[htb]
\begin{center}
\includegraphics[width=0.4\textwidth]{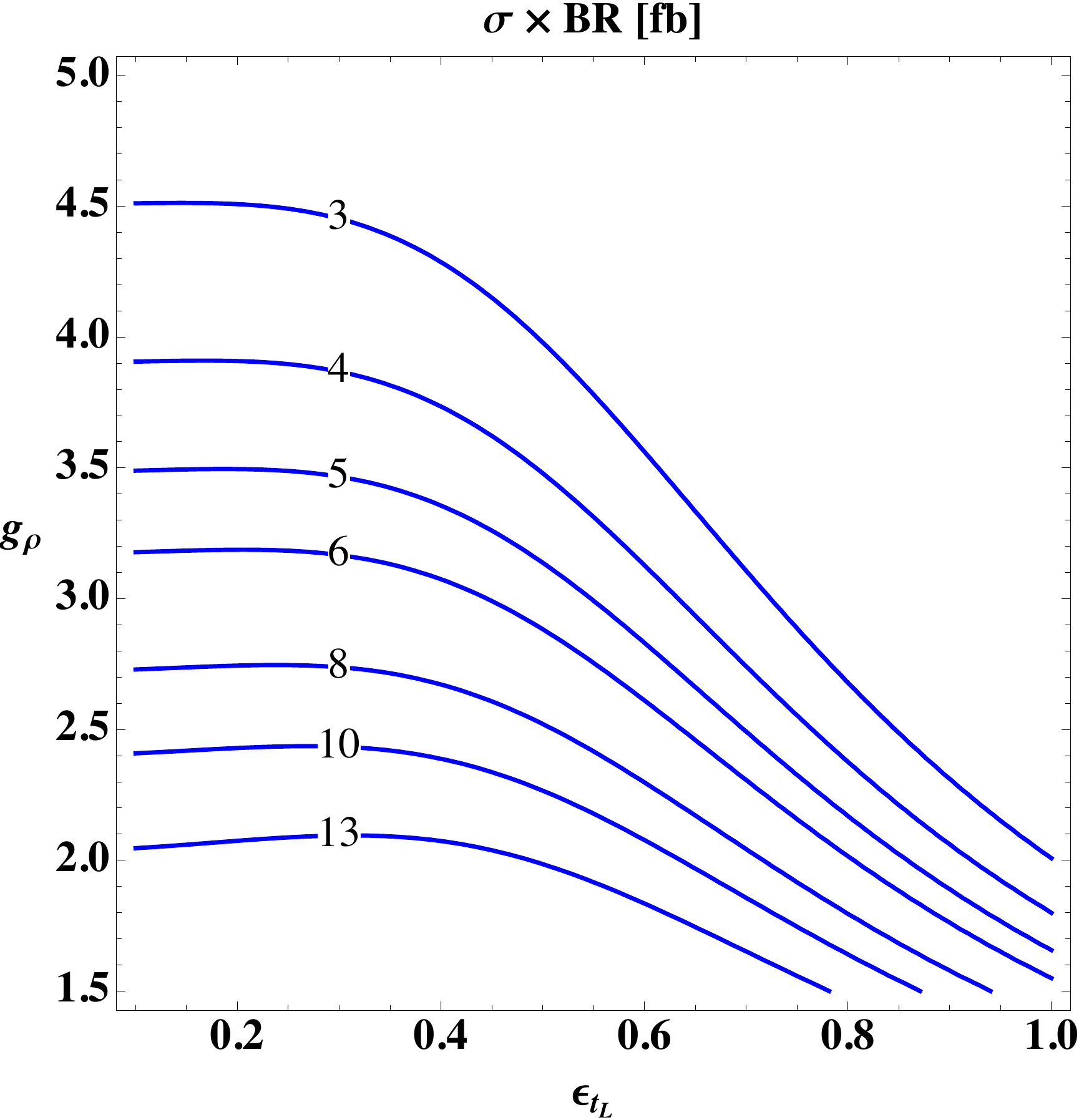}
\end{center}
\caption{Contours for the  $\sigma(pp\rightarrow\rho^\pm)\times BR(\rho^+ \rightarrow W^+ Z)$ .}
\label{fig:mgrho}
\end{figure} 

To begin with, we first consider the $\rho^\pm \rightarrow W^\pm Z$ channel for $M_{\rho^+} = $ 2 TeV to get a rough idea about the size of our parameters. The production cross sections of $\rho^\pm$ at 8 TeV LHC are given by,
\bea
\begin{split}
&\sigma(pp \rightarrow \rho^+)\sim   \frac{g^4}{g_\rho^2}\times 528 \,  \text{fb}, \\
&\sigma(pp \rightarrow \rho^-) \sim \frac{g^4}{g_\rho^2} \times  132 \,  \text{fb}
\end{split}
\eea
where we have applied a K factor $K = 1.34$~(\cite{Cao:2012ng}).
Remember for the dijet search, we cannot distinguish the $\rho^+$ from $\rho^-$, so that we should add them together.  The WZ excess being observed at ATLAS can be expressed as the production cross section of $\rho^\pm$ times the branching ratio of WZ, which can be expressed as a function of $ (g_\rho,\epsilon_{t_L}) $:
\bea
\begin{split}
&(\sigma(pp \rightarrow \rho^+)  + \sigma(pp \rightarrow \rho^-) )\times \text{BR}(\rho^+ \rightarrow W^+ Z)\\
& \sim 282 \, \text{fb} \times \frac{ \epsilon_g^2(1  + \epsilon_g^2)^2 }{ (1  + \epsilon_g^2)^2 +  1 + 48 \epsilon_g^4 + 12 \epsilon_{t_L}^4 - 24 \epsilon_g^2 \epsilon_{t_L}^2 } \\
\end{split}
\eea
where we have defined the $\epsilon_g = g/g_\rho,\epsilon_{t_L} $, which measure  the degree of compositeness for the SM gauge bosons. Note that we also neglected the $\xi$ correction in the calculation of the decay width. The behavior of $\sigma(pp \rightarrow \rho^\pm)\times\text{BR}(\rho^+ \rightarrow W^+ Z)$ as a function of $g_\rho$ and $\epsilon_{t_L}$  is shown in Fig.~\ref{fig:mgrho}, from which we can infer that a large regions $g_\rho \lesssim 4.0, \epsilon_{t_L} \lesssim 1.0 $ are allowed for the excess. Note that a not too small degree of compositeness  for the $t_L$ is needed to get the correct EWSB.

\begin{table}[t]
\begin{center}
\begin{tabular}{|lcr|c|c|c|c|}
\hline
\multicolumn{3}{|c|}{Channel} & Process&1.8 TeV &  2.0 TeV\\
\hline
ATLAS& $\ell\ell$&\cite{Aad:2014cka}& $pp\to Z'\to \ell \ell$&0.23 fb&  0.20 fb\\
ATLAS& $\ell\nu$&\cite{ATLAS:2014wra}& $pp\to W'\to \ell\nu$&0.54 fb& 0.44 fb\\
CMS &$\ell\ell$&\cite{Khachatryan:2014fba}& $pp\to Z'\to \ell \ell$ &0.24 fb& 0.24 fb\\
CMS &$\ell\nu$&\cite{Khachatryan:2014tva}& $pp\to W'\to \ell\nu$ &0.4 fb& 0.30 fb\\
\hline
\end{tabular}
\end{center}
\caption{Summary of  the relevant stringent observed bounds at 95 \% C.L. from LHC.}
\label{bounds}
\end{table}  

\begin{figure}[ht]
\begin{center}
\includegraphics[width=0.4\textwidth]{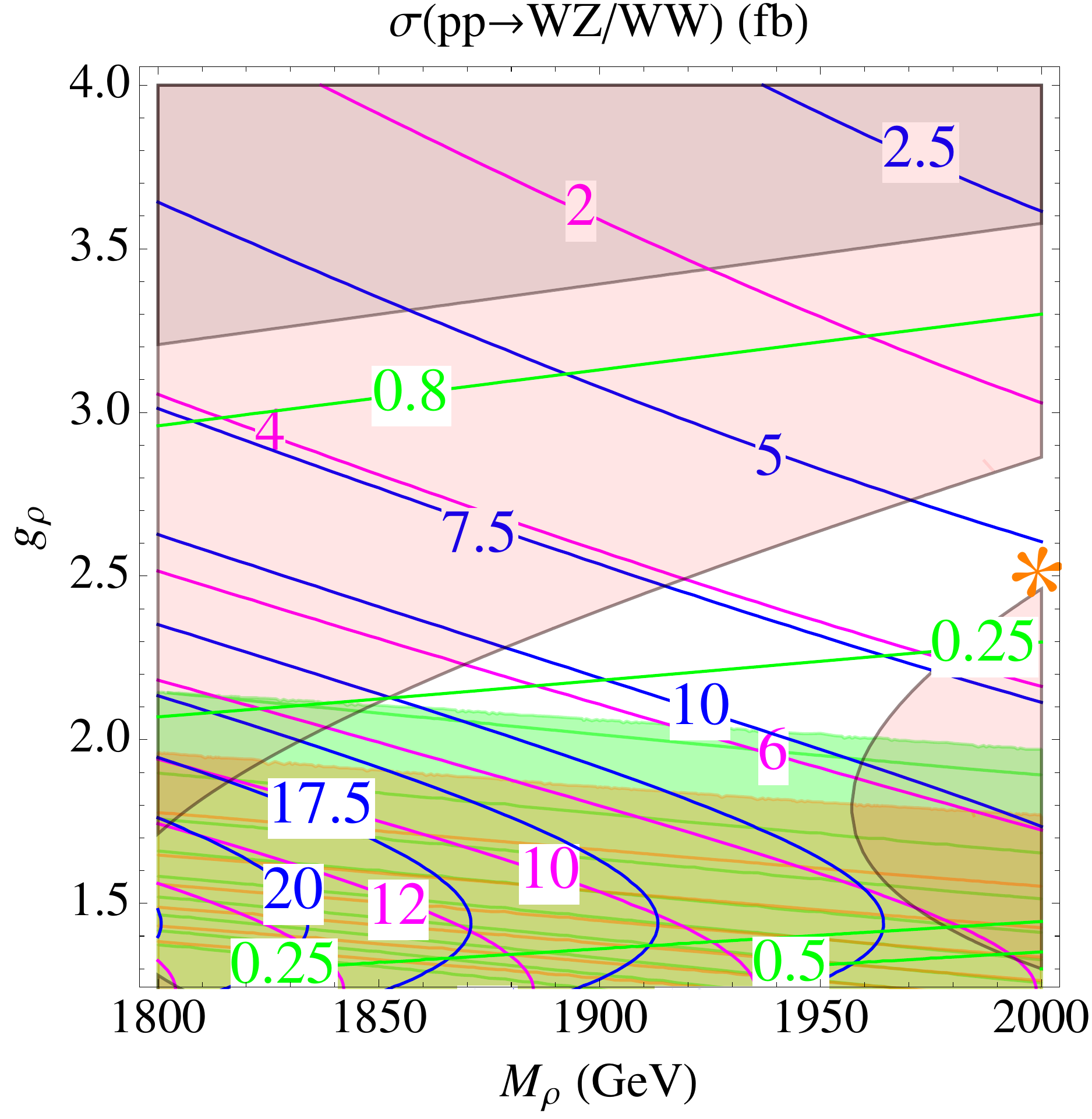}\\
\includegraphics[width=0.4\textwidth]{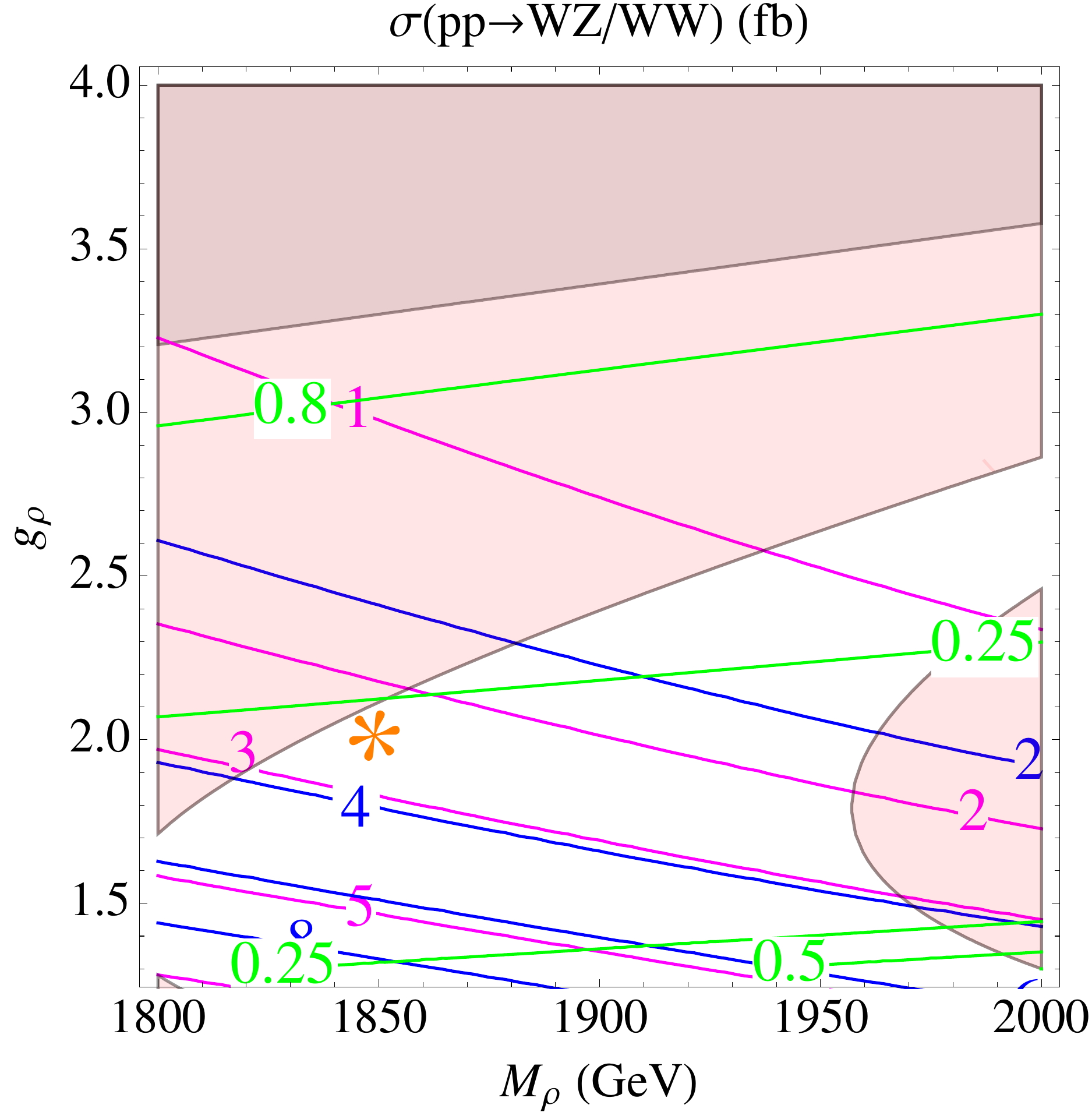}
\end{center}
\caption{Bounds from LHC searches and EWPT with $ \epsilon_{t_L} $ being fixed as 0.5 and 0.9 for the upper and downer pannel. For $\epsilon_{t_L} = 0.9$, we have assumed the compositness of lepton make the couplings with $\rho_L$ vanish and the bounds from leptonical searches disappear. The gray region is excluded by the theoretical consideration of the first Weinberg sum rule (see text).  The  blue/magenta contours corresponds to the cross section $\sigma(pp\rightarrow WZ/WW)$ and the light orange and green shaded region are excluded by  the LHC search $\sigma(pp\rightarrow Z^\prime \rightarrow e^+e^-/\mu^+\mu^-) $ (\cite{Aad:2014cka}) and 
$\sigma(pp\rightarrow W^\prime \rightarrow e\nu/\mu\nu)$ (\cite{Khachatryan:2014tva}).  The light red region indicates the constraints from EWPT at $95\%$ C.L.  and the green contours depict the ratio $\Gamma(h\rightarrow Z\gamma)/\Gamma(h\rightarrow Z\gamma)_{SM}.$ }
\label{fig:LHCbound}
\end{figure}

It is essential to check whether our resonance is consistent with the other $W^\prime, Z^\prime$ searches at LHC and the strongest constraints come from the dilepton searches (see Table.~\ref{bounds}). In Fig.~\ref{fig:LHCbound}, we plot  the three bounds in the $(M_\rho,g_\rho)$ plane  coming from the searches described above by choosing the following parameters:
\bea
\xi = 0.1,~ \alpha = 2.1, ~ \Delta = 1, ~ \epsilon_{t_L} = 0.5, 0.9 ,
\label{eq:xialpha}
\eea
where the mass ranges from 1.8 TeV to 2.0 TeV. We also show the bound (the grey region) from the first Weinberg sum rule:
\bea
f_\rho > \frac{f}{\sqrt{2}} \Rightarrow g_\rho < \frac{\sqrt{2} m_\rho}{f} 
\eea
The most strong contraints come from the fully leptonical search for the $W^\prime$, which gives $g_\rho \gtrsim 2.0$ for $M_\rho = 2$ TeV.
The white region with star labelled indicates the preferred range from EWPT at 95\%.  Combing all the constraints, we can see that the prefered vaule for $g_{\rho}$  is $ \sim 2- 2.5$
and the cross sections for the $\sigma$(WZ) and $\sigma$(WW) are
 consistent with  ATLAS di-boson excess.  For example, for $M_\rho \sim  2 $ TeV, $g_\rho \sim $ 2.5, $\epsilon_{t_L} = 0.5$, $\sigma( W Z) \sim $ 6.78 fb and $\sigma( W^+ W^-) \sim $ 3.74 fb. $(m_a, g_a)$ are obtained by solving the (eq.~\ref{SRI}, \ref{SRIInew}), where in this benchmark point, we get $(m_a,g_a) = (1.46 \text{TeV}, 2.68)$. Decreasing the mass and the coupling a little bit can make the cross section larger by a factor of $\sim$ 2 (see Table.~\ref{tab:benchmark}).

\begin{table}[h]
\begin{center}
\begin{tabular}{|c|c|c|c|c|c|}
	\hline
$(M_\rho[\text{TeV}], g_\rho)$ & $\sigma (WZ)$ &$\sigma (WW)$ & $m_a$ [TeV]& $g_a $&  $\frac{\Gamma(h\rightarrow Z\gamma)}{\Gamma(h\rightarrow Z\gamma)_{SM}}$\\
\hline
(2.0, 2.5)&  6.78 & 3.74   & 1.46 &2.68 &0.38 \\
\hline
(1.85, 2.0   )&4.23  & 3.32 &  1.29& 1.87 & 0.17\\
\hline
\end{tabular}
\end{center}
\caption{Benchmark points (star in Fig.~\ref{fig:LHCbound}) for  the  allowed region  assuming $\epsilon_{t_L} = 0.5$ and $\epsilon_{t_L} = 0.9$ for the first and second row. The cross sections are shown in fb.}
\label{tab:benchmark}
\end{table} 

\section{$h\rightarrow Z \gamma$ decay width}

The additional axial resonance is introduced to play the important role in relieving the stringent constraint of the S parameter
as discussed before. While, 
the additional non-diagonal gauge interactions from mixing effects between electroweak gauge bosons and 
the axial resonance
will contribute to a gauge invariant amplitude for the $h\rightarrow Z\gamma$ process.
In principle the $\rho_L$ will also contribute to the $h\rightarrow Z \gamma$, but its result will be  suppressed by  an additional $\xi$ or  $(g^2/g_\rho^2)$ compared with the axial resoance and thus can be safely neglected.

The general results for the $h \rightarrow Z \gamma$  decay width with respect to  the SM value can be expressed as a simple form:

\beq
\label{eq:hza}
\begin{split}
\frac{\Gamma(h\rightarrow Z \gamma)}{\Gamma(h\rightarrow Z \gamma)_{SM}} & \simeq
|a_W + 0.08 c_{Z\gamma}|^2\; , 
\end{split}
\eeq
where we have neglected the small top contribution and the coefficient $c_{Z\gamma}$ (see appendix for the definition of the coefficients.) can be calcualted  at 1-loop level (\cite{Cai:2013kpa}): 
\bea
 c_{Z \gamma} & = &  2\, c_{a W} \, c_{Z a W} \, c_{Z \gamma}^{(1)} \, (m_W, m_a)\; ,
\eea
with 
\bea
c_{Z \gamma}^{(1)} (m_W, m_a) &\simeq &  \cot \theta_w  \cdot \bigg(\frac{ 7 m_a^2}{4 m_W^2} + \frac{9}{2} \log \left(\frac{m_a^2}{m_W^2} \right)  \nonumber  \\ & & +  \frac{5 m_h^2-47 m_Z^2 -45 m_W^2}{36 m_W^2}  \bigg)   \nonumber \\  &+& \mathcal{O} \left(\frac{1}{m_a^2} \log \left(\frac{m_a^2}{m_W^2}\right) \right)\nonumber \\
\eea
The  couplings  in the mass eigenstates can be obtaind by diagonalizing the mixing mass matrix in the charged and neutral sector:
\beq
\begin{split}
a_W &= \sqrt{1 -  \xi} \left(1 + \mathcal{O}(\frac{g^2}{g_a^2}\xi)\right) \\
c_{a W}  &=  - \frac{\Delta}{\sqrt{2}} \frac{g }{g_a}  \sqrt{\xi}\sqrt{1 - \xi} \left(1 + \mathcal{O}(\frac{g^2}{g_a^2}\xi)\right)\\
c_a &= \frac{\Delta^2 g^2 }{4g_a^2}  \xi \sqrt{1-\xi}\left(1 + \mathcal{O}(\frac{g^2}{g_a^2}\xi)\right)\\
c_{ZaW} &= (1 - \frac{1}{2}\frac{1}{\cos^2\theta_W}) \frac{\Delta}{\sqrt{2}}\frac{g}{g_a}\sqrt{\xi}\\
\end{split}
\eeq

Note that we find different results with Ref.~\cite{Cai:2013kpa}, especially for the couplings $a_W, c_{ZaW}$, where  Ref.~\cite{Cai:2013kpa}  has missed one term for the $c_{ZaW}$. The non-diagonal contribution interferes with SM amplitude deconstructively, as  shown in Fig.~\ref{fig:LHCbound}, where we plot the contours  (in green) for the $\Gamma(h\rightarrow Z \gamma)/\Gamma(h\rightarrow Z \gamma)_{SM}$ with the same parameters in eq.~\ref{eq:xialpha}. From the benchmark points shown in Table.~\ref{tab:benchmark}, the deviation from the SM can be as large as 70\% and this could provide the strong evidence on the existence of axial resonances if such a deviation is observed at 14 TeV LHC.



\section{Conclusion and Outlook}

In this paper, we have considered the possibility of spin one resonances in the minimal composite Higgs scenario based on the $SO(5)/SO(4) \times U(1)_X$ ($SO(4) \simeq SU(2)_L \times SU(2)_R $) symmetry breaking pattern to account for the recent diboson excess reported in both ATLAS and CMS Collaborations. The relevant model is constructed in the CCWZ language with vector meson dominance and imposing the first Weinberg sum rule and positive SM gauge boson form factors. We emphasis that the presence of axial vector resonance $a$ as a bi-doublet of $SU(2)_L \times SU(2)_R$ is crucial to dramatically improve the EWPT if a $SU(2)_L$ triplet vector boson $\rho$ with a mass $1.8 \sim 2$ TeV is used to explain the diboson excess. The suppressed couplings between $a$ and SM quarks ($\sim \sqrt{\xi} g / g_a$) suggest that $a$ is hided in the 7 $\sim$ 8 TeV LHC resonance searches. We find that our model can fit the diboson excess very well if the left-handed SM light quarks, charged leptons and top/bottoms have zero, zero or moderately small and moderate or large composite components for reasonable value of $g_\rho$ and $f$. If such a diboson excess is indeed confirmed at the $13 \sim 14$ TeV LHC run two, the large deviation in the $h\rightarrow Z \gamma$ decay width would be a strong hint for the existence of $a$ expected in our model.

There are indeed various interesting aspects to pursue since current paper only provide the first order sketch of our model. Here we just listed a few of them below: (1): unitarity bounds on the cut off scale from the longitudinal gauge boson scattering by including both $\rho$ and $a$. (2): realistic log divergent Coleman Weinberg Higgs potential from a composite top, $\rho$ and $a$ to get the correct Higgs mass (3): direct axial vector boson searches at the $13 \sim 14$ TeV LHC Run II (4): full calculation of $h \rightarrow \gamma \gamma$ and $h \rightarrow Z \gamma$ including both $\rho$ and $a$, etc. 
We will leave the above issues for future studies and expect that our scenario can attract people's interests in the axial vector bosons studies. 

\section{Appendix A:  Decay width and triple couplings}

In the mass eigenbasis, the general Lagrangian involving cubic interactions terms between the heavy resonances and the SM fields reads (\cite{Greco:2014aza}):
\begin{equation}
\label{eq:cubic}
\begin{split}
\mathcal{L}_\rho   =  \, 
& i  g_{\rho^+ W Z} \, \big[  (\partial_\mu \rho_{\nu}^+ - \partial_\nu \rho_{\mu}^+ ) W^{\mu -}Z^{\nu}    \\
  &                          -(\partial_\mu W_\nu^- - \partial_\nu W_\mu^- ) \rho^{\mu +} Z^{\nu } \phantom{ig_{\rho^+ W Z}\,\big[}       \\    
 &                          +  (\partial_\mu Z_\nu - \partial_\nu Z_\mu ) \rho^{\mu +}W^{\nu -}  + h.c.\big]  \\[0.15cm]
& + i g_{\rho^0 W W}\, \big[  (\partial_\mu W_\nu^+ - \partial_\nu W_\mu^+ ) W^{\mu -}\rho^{0\nu}  \\[0.15cm]
&                                  + \frac 12 (\partial_\mu \rho_{\nu}^0 - \partial_\nu \rho_{\mu}^0 ) W^{\mu +}W^{\nu -}  + h.c.\big]  \\[0.15cm]
& + g_{\rho^+ W h}  \, ( h \rho_{\mu}^+ W^{\mu-} + h.c.)  +  g_{\rho^0 Z h}  \, h  \rho_{\mu}^0 Z^{\mu}   \\[0.15cm]
& + \frac{1}{\sqrt{2}} \, g_{\rho^+ ff^\prime} \left( \rho^+_{\mu}\bar{f} \gamma^\mu  f ^\prime + h.c. \right) \\[0.15cm]
& + \, g_{\rho^+ ff} \left( \rho^+_{\mu}\bar{f} \gamma^\mu  f  + h.c. \right) \\[0.15cm]
\end{split}
\end{equation}
where  $f$ stands for any of the SM chiral fermions and $\rho$ can be either $\rho_L$ or $\rho_R$. We will only show the couplings  in the large coupling limit $g_\rho \gg g$ for $\rho_L$ and keep the leading term in $\xi$. We will make some comments for $\rho_R$ in the end.
First, we list the mass formulae for the $\rho_L$:
\beq
M^2_{\rho^{\pm,0}} \sim m^2_{\rho} \left(1 + \frac{g^2}{g_{\rho}^2}\right)\, ,
\eeq
Secondly, we show the couplings involving the SM gauge bosons and the Higgs:
\bea
&&g_{\rho^+ WZ} \sim \frac{m_Z m_W}{m_\rho^2} g_{\rho},~ g_{\rho^+ Wh} \sim m_W g_{\rho}\; ,\nonumber\\
&&g_{\rho^0 WW} \sim \frac{ m_W^2}{m_\rho^2} g_{\rho},~g_{\rho^0 Zh} \sim m_Z g_{\rho}\, .
\eea
For the couplings with fermions, there are in general two sources:  the mixing of the SM gauge boson with composite $\rho$ resonances and the mixing of the  SM fermions with composite fermionic resonances. The first effect is universal and scales like $g^2/g_\rho$. The second effect depends on the masses of the fermion and only the SM top can have a significant size. Under the assumption of  $t_R$ being  a $SO(4)$ singlet and the coupling involving it will start at $\xi$.  So only the partial compositeness of $t_L$ has an important impact on the phenomenology of the $\rho$ and we define $\epsilon_{t_L}$ as the degree of its compositeness. We summarize the results below:
\bea
&&g_{\rho^+f_{el}f_{el}^\prime}\sim  - \frac{g^2}{g_\rho},  ~ g_{\rho^+ t_L b_L} \sim  -\frac{g^2}{g_\rho} + g_\rho \epsilon_{t_L}^2\;,\nonumber\\
&& g_{\rho^0f_{el}f_{el}}  \sim - T^{3L}_f \frac{g^2}{g_\rho},\nonumber\\
&&g_{\rho^0 t_L t_L}  \sim \frac{1}{2} (- \frac{g^2}{g_\rho} + g_\rho \epsilon_{t_L}^2) \;, \\
&&g_{\rho^0 b_L b_L}  \sim -\frac{1}{2} (-\frac{g^2}{g_\rho} + g_\rho \epsilon_{t_L}^2) \;, 
\eea
where we have neglected the $\xi$ correction and $f_{el}$ is the fully elementary chiral  SM fermions. The difference of $\rho_R$ is that it only mixing with the SM hyper-charge field, as a result only $\rho_R^0$ has a non-zero coupling with SM elementary fermions before EWSB and scale like $- Y g^{\prime 2}/g_\rho$, where $Y$ is the hyper charge field.
We finally present the analytical formulae for the decay widths of $\rho$ :
\bea
&&\Gamma(\rho^+\rightarrow W^+ Z)=M^5_{\rho^+}g^2_{\rho^+ WZ}/(192\pi m_W^2 m_Z^2),~\nonumber \\
&&\Gamma(\rho^+\rightarrow W^+ h)=M_{\rho^+}g^2_{\rho^+ Wh}/(192\pi m_W^2),~\nonumber\\
&&\Gamma(\rho^+\rightarrow \psi_f \bar{\psi}_{f^\prime}) =N_c M_{\rho^+}g^2_{\rho^+ ff^\prime} /48\pi, \nonumber\\
&&\Gamma(\rho^0\rightarrow W^+ W^-)=M^5_{\rho^0}g^2_{\rho^0 WW}/(192\pi  m_W^4),~\nonumber\\
&&\Gamma(\rho^0\rightarrow Z h)=M_{\rho^0} g^2_{\rho^0 Zh}/(192\pi m_Z^2),~\nonumber\\
&&\Gamma(\rho^0\rightarrow \psi_f \bar{\psi}_f ) =N_c M_{\rho^0}g^2_{\rho^0 ff}/24\pi.
\eea

where $N_c$ is the color factor for the SM fermions  and $f$ denotes any SM chiral fermions.

\section{Appendix B: The model and the couplings relevant to $h\rightarrow Z\gamma$ }
\label{hzac}
The effective Lagrangian parametrizing the Higgs interactions with the gauge bosons is:
\bea
 \mathcal{L}_{eff} &=& a_W \frac{2m_W^2}{v} h W_\mu ^+ W_\mu
^- + a_Z \frac{m_Z^2}{v} h Z_\mu  Z_\mu  \\
&&+ c_\rho \frac{2
m_\rho^2}{v} h \rho_{L \mu}^+ \rho_{L \mu}^-  \nonumber  \\ 
& & + c_a \frac{2
m_a^2}{v}   h a_{ \mu}^+ a_{\mu}^-   +
 c_{a W}  \frac{ m_a^2}{v} h \left( W_\mu^+ a_{\mu}^- +W_\mu^- a_{\mu}^+\right) \nonumber  \\ & & +  c_{a \rho_L}  \frac{ m_a^2}{v}  h \left( \rho_{L \mu}^+ a_{\mu}^- +\rho_{L \mu}^- a_{\mu}^+\right) +  {c_f}  \left(
\frac{m_f}{v}\bar f f \right) h   \nonumber  \\ & & +  c_\gamma ~ \frac{\alpha }{8 \pi
v } h A^{\mu \nu } A_{\mu \nu }  + c_{ Z\gamma} \frac{\alpha }{4
\pi v }  h Z^{\mu \nu } A_{\mu \nu }\; ,
 \eea
We can obtain the relevant  couplings for $h\rightarrow Z \gamma$ process in the presence of the axial resonance by diagonalizing the mass matrix at leading order:
\beq
\begin{split}
a_W &= \sqrt{1 -  \xi} \left(1 + \mathcal{O}(\frac{g^2}{g_a^2}\xi)\right) \\
a_Z &=\sqrt{1 -  \xi} \left(1 + \mathcal{O}(\frac{g^2}{g_a^2}\xi)\right)   \\
c_{a W}  &=  - \frac{\Delta}{\sqrt{2}} \frac{g }{g_a}  \sqrt{\xi}\sqrt{1 - \xi} \left(1 + \mathcal{O}(\frac{g^2}{g_a^2}\xi)\right)\\
c_a &= \frac{\Delta^2 g^2 }{4g_a^2}  \xi \sqrt{1-\xi}\left(1 + \mathcal{O}(\frac{g^2}{g_a^2}\xi)\right)\\
c_{ZaW} &= (1 - \frac{1}{2}\frac{1}{\cos^2\theta_W}) \frac{\Delta}{\sqrt{2}}\frac{g}{g_a}\sqrt{\xi}\\
c_a &=\mathcal{O}( \frac{g^2}{g_a^2}\xi )\\
\end{split}
\eeq
By power counting, we can roughly estimate the other couplings involving the $\rho_L$:
\beq
\begin{split}
c_\rho &\sim \frac{g^2}{g_\rho^2}\xi ,\qquad c_{a\rho_L} \sim \frac{g}{g_\rho}   \frac{g}{g_a} \sqrt{\xi} \\
c_{Za\rho_L} &\sim \frac{g}{g_\rho}   \frac{g}{g_a} \sqrt{\xi} \\
\end{split}
\eeq
As expected, the mixing couplings for the $(a,\rho_L)$ are suppressed by $g/g_\rho$ .

\appendix

\vspace*{-15pt}



\begin{thebibliography}{99}


\bibitem{Aad:2015owa}
  G.~Aad {\it et al.}  [ATLAS Collaboration],
  ``Search for high-mass diboson resonances with boson-tagged jets in proton-proton collisions at $\sqrt{s}$ = 8 TeV with the ATLAS detector,''
  arXiv:1506.00962 [hep-ex].
  
  \bibitem{Khachatryan:2014hpa} 
  V.~Khachatryan {\it et al.} [CMS Collaboration],
  JHEP {\bf 1408}, 173 (2014)
  [arXiv:1405.1994 [hep-ex]].
\bibitem{Khachatryan:2014gha} 
  V.~Khachatryan {\it et al.} [CMS Collaboration],
  JHEP {\bf 1408}, 174 (2014)
  [arXiv:1405.3447 [hep-ex]].
  
\bibitem{Khachatryan:2015sja} 
  V.~Khachatryan {\it et al.} [CMS Collaboration],
  Phys.\ Rev.\ D {\bf 91}, no. 5, 052009 (2015)
  [arXiv:1501.04198 [hep-ex]].
 \bibitem{Cacciapaglia:2015nga} 
  G.~Cacciapaglia, A.~Deandrea and M.~Hashimoto,
  arXiv:1507.03098 [hep-ph].
\bibitem{Carmona:2015xaa} 
  A.~Carmona, A.~Delgado, M.~Quiros and J.~Santiago,
  arXiv:1507.01914 [hep-ph].
\bibitem{Gao:2015irw} 
  Y.~Gao, T.~Ghosh, K.~Sinha and J.~H.~Yu,
  arXiv:1506.07511 [hep-ph].
  
\bibitem{Cao:2015lia} 
  Q.~H.~Cao, B.~Yan and D.~M.~Zhang,
  arXiv:1507.00268 [hep-ph].

\bibitem{Cheung:2015nha} 
  K.~Cheung, W.~Y.~Keung, P.~Y.~Tseng and T.~C.~Yuan,
  arXiv:1506.06064 [hep-ph].
\bibitem{Abe:2015uaa} 
  T.~Abe, T.~Kitahara and M.~M.~Nojiri,
  arXiv:1507.01681 [hep-ph].
\bibitem{Sanz:2015zha} 
  V.~Sanz,
  arXiv:1507.03553 [hep-ph].
\bibitem{Thamm:2015csa} 
  A.~Thamm, R.~Torre and A.~Wulzer,
  arXiv:1506.08688 [hep-ph].
  
\bibitem{Anchordoqui:2015uea} 
  L.~A.~Anchordoqui, I.~Antoniadis, H.~Goldberg, X.~Huang, D.~Lust and T.~R.~Taylor,
  arXiv:1507.05299 [hep-ph].
\bibitem{Omura:2015nwa} 
  Y.~Omura, K.~Tobe and K.~Tsumura,
  arXiv:1507.05028 [hep-ph].
\bibitem{Chen:2015xql} 
  C.~H.~Chen and T.~Nomura,
  arXiv:1507.04431 [hep-ph].
\bibitem{Chiang:2015lqa} 
  C.~W.~Chiang, H.~Fukuda, K.~Harigaya, M.~Ibe and T.~T.~Yanagida,
  arXiv:1507.02483 [hep-ph].
\bibitem{Dobrescu:2015yba} 
  B.~A.~Dobrescu and Z.~Liu,
  arXiv:1507.01923 [hep-ph].

\bibitem{Brehmer:2015cia} 
  J.~Brehmer, J.~Hewett, J.~Kopp, T.~Rizzo and J.~Tattersall,
  arXiv:1507.00013 [hep-ph].
\bibitem{Dobrescu:2015qna} 
  B.~A.~Dobrescu and Z.~Liu,
  arXiv:1506.06736 [hep-ph].
\bibitem{Hisano:2015gna} 
  J.~Hisano, N.~Nagata and Y.~Omura,
  arXiv:1506.03931 [hep-ph].
\bibitem{Chao:2015eea} 
  W.~Chao,
  arXiv:1507.05310 [hep-ph].
  
  
\bibitem{Orgogozo:2012ct} 
  A.~Orgogozo and S.~Rychkov,
  JHEP {\bf 1306}, 014 (2013);
  A.~Orgogozo and S.~Rychkov,
  JHEP {\bf 1203}, 046 (2012)
  [arXiv:1111.3534 [hep-ph]].

  
  
\bibitem{Contino:2015mha} 
  R.~Contino and M.~Salvarezza,
  arXiv:1504.02750 [hep-ph].

\bibitem{Barbieri:2007bh} 
  R.~Barbieri, B.~Bellazzini, V.~S.~Rychkov and A.~Varagnolo,
  Phys.\ Rev.\ D {\bf 76}, 115008 (2007)
  [arXiv:0706.0432 [hep-ph]].
  
\bibitem{Contino:2011np} 
  R.~Contino, D.~Marzocca, D.~Pappadopulo and R.~Rattazzi,
  JHEP {\bf 1110}, 081 (2011)
  [arXiv:1109.1570 [hep-ph]].

  
\bibitem{Marzocca:2012zn} 
  D.~Marzocca, M.~Serone and J.~Shu,
  JHEP {\bf 1208}, 013 (2012)
  [arXiv:1205.0770 [hep-ph]].



\bibitem{Witten:1983ut} 
  E.~Witten,
  Phys.\ Rev.\ Lett.\  {\bf 51}, 2351 (1983).

\bibitem{CCWZ}
S. R. Coleman, J. Wess and B. Zumino,  Phys. Rev. 177, 2239 (1969); C. G. Callan Jr., S. R. Coleman, J. Wess and B. Zumino,  Phys. Rev. 177, 2247 (1969).



\bibitem{Marzocca}
    D. Marzocca,   M. Serone and  J. Shu,
    JHEP 1208 (2012) 013
    [arXiv:1205.0770 [hep-ph]].

  

\bibitem{CMSWZhad_2014} 
  [CMS Collaboration],
  ``Search for massive resonances in dijet systems containing jets tagged as W or Z boson decays in pp collisions at $\sqrt{s}$ = 8 TeV''  [\href{http://arxiv.org/abs/1405.1994}arXiv:1405.1994 [hep-ex]].




\bibitem{Aad:2014cka} 
  G.~Aad {\it et al.} [ATLAS Collaboration],
  Phys.\ Rev.\ D {\bf 90}, no. 5, 052005 (2014)
  [arXiv:1405.4123 [hep-ex]].
\bibitem{Weinberg:1967kj}
  S.~Weinberg,
  Phys.\ Rev.\ Lett.\  {\bf 18} (1967) 507.
\bibitem{Khachatryan:2014fba} 
  V.~Khachatryan {\it et al.} [CMS Collaboration],
  JHEP {\bf 1504}, 025 (2015)
  [arXiv:1412.6302 [hep-ex]].
\bibitem{ATLAS:2014wra} 
  G.~Aad {\it et al.} [ATLAS Collaboration],
  JHEP {\bf 1409}, 037 (2014)
  [arXiv:1407.7494 [hep-ex]].
\bibitem{Khachatryan:2014tva} 
  V.~Khachatryan {\it et al.} [CMS Collaboration],
  Phys.\ Rev.\ D {\bf 91}, no. 9, 092005 (2015)
  [arXiv:1408.2745 [hep-ex]].

\bibitem{Khachatryan:2015bma} 
  V.~Khachatryan {\it et al.} [CMS Collaboration],
  arXiv:1506.01443 [hep-ex].
  

\bibitem{Khachatryan:2015sma} 
  V.~Khachatryan {\it et al.} [CMS Collaboration],
  arXiv:1506.03062 [hep-ex].

  
\bibitem{Cao:2012ng}
  Q.~H.~Cao, Z.~Li, J.~H.~Yu and C.~P.~Yuan,
  Phys.\ Rev.\ D {\bf 86} (2012) 095010
  [arXiv:1205.3769 [hep-ph]].

\bibitem{Cai:2013kpa}
  H.~Cai,
  JHEP {\bf 1404} (2014) 052
  [arXiv:1306.3922 [hep-ph]].

    

\bibitem{Greco:2014aza}
  D.~Greco and D.~Liu,
  JHEP {\bf 1412} (2014) 126
  [arXiv:1410.2883 [hep-ph]].
\bibitem{Bian:2015zha} 
  L.~Bian, J.~Shu and Y.~Zhang,
  arXiv:1507.02238 [hep-ph].
  
  \bibitem{Grojean}
  C.~Grojean, W.~Skiba and J.~Terning,
  Phys.\ Rev.\ D {\bf 73}, 075008 (2006)
  [hep-ph/0602154].
  
\bibitem{Agashe:2003zs} 
  K.~Agashe, A.~Delgado, M.~J.~May and R.~Sundrum,
  JHEP {\bf 0308}, 050 (2003)
  [hep-ph/0308036].

\bibitem{Shu:2011au} 
  J.~Shu, K.~Wang and G.~Zhu,
  Phys.\ Rev.\ D {\bf 85}, 034008 (2012)
  [arXiv:1104.0083 [hep-ph]].
\bibitem{Csaki:2003zu} 
  C.~Csaki, C.~Grojean, L.~Pilo and J.~Terning,
  Phys.\ Rev.\ Lett.\  {\bf 92}, 101802 (2004)
  [hep-ph/0308038].
\bibitem{Cacciapaglia:2004rb} 
  G.~Cacciapaglia, C.~Csaki, C.~Grojean and J.~Terning,
  Phys.\ Rev.\ D {\bf 71}, 035015 (2005)
  [hep-ph/0409126].
            

    
\end{thebibliography}
\end{document}